\documentclass[]{spie}  %>>> use for US letter paper
\usepackage{amsmath,amsfonts,amssymb}
\usepackage{graphicx}
\usepackage[colorlinks=true, allcolors=blue]{hyperref}
\usepackage{physics}
\usepackage{array,multirow,boldline}
\newcolumntype{C}[1]{>{\centering\let\newline\\\arraybackslash\hspace{0pt}}m{#1}} % Cell width

\title{High-fidelity quantum tomography with imperfect measurements}

\author[a,b]{B. I. Bantysh}
\author[a,b]{D. V. Fastovets}
\author[a,b,c]{Yu. I. Bogdanov}
\affil[a]{Valiev Institute of Physics and Technology of Russian Academy of Sciences, Moscow, Russia}
\affil[b]{National Research University of Electronic Technology (MIET), Moscow, Russia}
\affil[c]{National Research Nuclear University (MEPhI), Moscow, Russia}

\authorinfo{E-mail: bbantysh60000@gmail.com}

\begin{document} 
\maketitle

\begin{abstract}
In the current work we address the problem of quantum process tomography (QPT) in the case of imperfect preparation and measurement of the states which are used for QPT. The fuzzy measurements approach which helps us to efficiently take these imperfections into account is considered. However, to implement such a procedure one should have a detailed information about the errors. An approach for obtaining the partial information about them is proposed. It is based on the tomography of the ideal identity gate. This gate could be implemented by performing the measurement right after the initial state preparation. By using the result of the identity gate tomography we were able to significantly improve further QPT procedures. The proposed approach has been tested experimentally on the IBM superconducting quantum processor. As a result, we have obtained an increase in fidelity from 89\% to 98\% for Hadamard transformation and from 77\% to 95\% for CNOT gate.
\end{abstract}


\section{INTRODUCTION}\label{sect:intro}

During the development of quantum information technologies the problem of determining the quality of element base of quantum computational devices is constantly arising. Nowadays, one of the most common approaches to this problem is the randomized benchmarking technique for experimental determination of quantum process fidelity by implementing the chains of random transformations\cite{knill2008,magesan2012}.

However, the fidelity determination is not enough if one needs to get a detailed gate characterization in order to localize the origination of its imperfections and make an effort to eliminate them. The most complete solution to this task is provided by \textit{quantum process tomography} (QPT)\cite{chuang1997,mohseni2008,bogdanov2013njp,bogdanov2012micro,altepeter2003,obrien2004,bialczak2010}. Due to \textit{Choi--Jamiolkowski isomorphism}, QPT could be reduced to the estimation of the specific quantum state (Choi--Jamiolkowski state) in the Hilbert space of a higher dimension\cite{holevo2013}. As such, the complete positivity of a quantum channel is provided by the positivity of this state density matrix. In Sect.~\ref{sect:qpt} we describe the basic QPT procedure that provides an asymptotically unbiased and statistically significant estimation of a Choi--Jamiolkowski state in the case of an ideal experiment.

In real experimental conditions the measurement setup is not ideal, which makes it difficult to perform a precise QPT. The presence of state preparation and measurement (SPAM) errors results in two significant issues. On the one hand, the quantum process estimation becomes inconsistent which limits the fidelity that could in principle be achieved in experiment. On the other hand, SPAM-errors introduce to the reconstruction result additional non-unitary processes, which the gate does not actually perform.

The most effective solution to this problem lies in the modification of the reconstruction procedure in order to include SPAM-errors in the model \cite{anis2012,bogdanov2016spie_fuzzy,bogdanov2018lp}. In Sect.~\ref{sect:fuzzy_measurements} we introduce a simple method for such a modification. However, to implement it one should have a detailed information about the processes which take place during state preparation and measurements. This information could be obtained by a proper physical description of the measurement setup. If this description could not be performed with a high accuracy one could obtain it experimentally. The starting point here is to measure an ideal identity gate by measuring the state right after its preparation\cite{merkel2013}. In Sect.~\ref{sect:empty_tomo} we demonstrate how these kinds of measurements let us estimate the quantum channel that partly describes the SPAM-errors. The result of this estimation could significantly improve the quality of further QPT.

The methods proposed in the current work have been implemented experimentally using the IBM superconducting quantum processor which is available through the cloud service\cite{ibm2017}. The results are presented in Sect.~\ref{sect:ibm}. In the considered example our approach resulted in an increase in fidelity from 89\% to 98\% for Hadamard transformation and from 77\% to 95\% for CNOT gate.

Finally, in Sect.~\ref{sect:conclusions} we briefly describe the main results of this work.

\section{QUANTUM PROCESS TOMOGRAPHY}\label{sect:qpt}

\subsection{Quantum operations formalism}\label{sect:operations}

Formally, an arbitrary quantum process on a Hilbert space $\mathcal{H}$ of dimension $s$ could be described as a completely positive map\cite{kraus1983,breuer2007,choi1975,nielsen2000}. Acting on arbitrary density matrix $\rho$, the process could be described in terms of operator sum:
\begin{equation}\label{eq:kraus_decomp}
  \mathcal{E}(\rho) = \sum_{k=1}^r{E_k \rho E_k^\dagger},
\end{equation}
where $E_k$ are Kraus operators. The process rank $r$ characterizes the amount of possible paths of state evolution. In order to preserve the trace of a density matrix Kraus operators should be subjected to the following normalization condition:
\begin{equation}\label{eq:kraus_normalization}
  \sum_{k=1}^r{E_k^\dagger E_k} = I_s,
\end{equation}
where $I_s$ is the identity matrix of dimension $s \times s$. The case $r = 1$ corresponds to a unitary evolution.

Let us consider a complete orthogonal set of operators composed of orthonormal set of vectors $\ket m$ ($m = 1, 2, \dots, s$):
\begin{equation}\label{eq:chi_basis}
  A_m \equiv A_{(m_1-1)s+m_2} = \ketbra{m_2}{m_1}, \quad m = 1,2,\dots,s^2, \quad m_1,m_2 = 1,2,\dots,s.
\end{equation}
Let us decompose each of $E_k$ using this basis:
\begin{equation}
  E_k = \sum_{m=1}^{s^2}{e_{mk}A_m}, \quad k = 1,2,\dots,r.
\end{equation}
The decomposition coefficients form the matrix $e$ of dimension $s^2 \times r$. One can then rewrite \eqref{eq:kraus_decomp} as
\begin{equation}
  \mathcal{E}(\rho) = \sum_{m,n=1}^{s^2}{\chi_{mn} A_m \rho A_n^\dagger},
\end{equation}
where we have introduced the $\chi$-matrix of the process $\mathcal{E}$:
\begin{equation}\label{eq:chi_def}
  \chi = ee^\dagger.
\end{equation}
This matrix is Hermitian and positive, and its trace is equal to $s$. Thus, any quantum transformation in Hilbert space of dimension $s$ could be completely described with a density matrix
\begin{equation}
  \rho_\chi = \frac1s \chi
\end{equation}
in the Hilbert space of dimension $s^2$ (Choi--Jamiolkowski isomorphism\cite{holevo2013}). As a result, QPT could be reduced to the tomography of the quantum state $\rho_\chi$. For example, if the process acts on the state of one qubit, one can define a two-qubit quantum state that corresponds to this process.

Let us consider the $\chi$-matrix as a joint state of subsystems $A$ and $B$, each of dimension $s$. Then the normalization condition \eqref{eq:kraus_normalization} could be expressed with the following equation:
\begin{equation}\label{eq:chi_normalization}
  \Tr_B(\chi) = \sum_{m=1}^s{ \qty(I_s \otimes \bra m)\chi\qty(I_s \otimes \ket m) } = I_s,
\end{equation}
where the summation is performed by an arbitrary orthonormal basis from $\mathcal{H}$.

\subsection{Accumulation of sample measurements for $\chi$-matrix}\label{sect:chi_measurement}

The statistical data for $\chi$-matrix measurements could be obtained in the following way\cite{bogdanov2013njp,bogdanov2012micro}. Consider a quantum state from the fixed set of state $\ket*{\psi^{(i)}}$ ($i = 1,2,\dots,m_p$) as the input of the process $\mathcal{E}$. The state $\mathcal{E}(\ketbra*{\psi^{(i)}})$ at the output is subjected to measurements with measurement operators $\Lambda_j$ ($j = 1,2,\dots,m_m$). According to the Born law,
\begin{equation}\label{eq:dm_prop}
  p_j( \mathcal{E}(\ketbra*{\psi^{(i)}}) ) = \Tr( \mathcal{E}(\ketbra*{\psi^{(i)}}) \Lambda_j )
\end{equation}
is the probability to get the $j$-th result. It can be shown that this probability is equal to the value that one obtains by measuring $\chi$-matrix of the process $\mathcal{E}$:
\begin{equation}\label{eq:chi_prob}
  p_j^{(i)}(\chi) = p_j( \mathcal{E}(\ketbra*{\psi^{(i)}}) ) = \Tr(\chi \Lambda_j^{(i)}), \quad
  \Lambda_j^{(i)} = \qty(\ketbra*{\psi^{(i)}})^\ast \otimes \Lambda_j.
\end{equation}
Here ``$\ast$'' stands for the complex conjugation. As $\chi$-matrix is normalized to $s$ the probabilities $p_j^{(i)}(\chi)$ are also normalized by $s$, but not by unity.

Due to \eqref{eq:chi_prob}, the statistical data over different $\ket*{\psi^{(i)}}$ and $\Lambda_j$ is equivalent to the data one would obtain by measuring the $\chi$-matrix directly. States $\ket*{\psi^{(i)}}$ and operators $\Lambda_j$ form the protocol of $\chi$-matrix measurements.

Below we will use the protocol with cube symmetry. For one-qubit gate estimation the input states are the eigenvectors of the standard Pauli matrices $\sigma_x$, $\sigma_y$ and $\sigma_z$ ($m_p = 6$):
\begin{equation}\label{eq:proto_prep}
  \ket*{\psi^{(1,2)}} = \frac1{\sqrt2}\mqty(1 \\ \pm 1), \quad
  \ket*{\psi^{(3,4)}} = \frac1{\sqrt2}\mqty(1 \\ \pm i), \quad
  \ket*{\psi^{(5)}} = \mqty(1 \\ 0), \quad
  \ket*{\psi^{(6)}} = \mqty(0 \\ 1).
\end{equation}
At the output the measurements of observables $\sigma_x$, $\sigma_y$ and $\sigma_z$ are performed (Pauli measurements). This gives us $m_m = 6$ measurement operators --- projectors to the states from the set \eqref{eq:proto_prep}:
\begin{equation}\label{eq:proto_meas}
  \Lambda_j = \ketbra*{\psi^{(j)}}, \quad  j = 1,2,\dots,6.
\end{equation}
Thus, the cube measurements protocol for $\chi$-matrix contains $m = m_p \cdot m_m = 36$ measurement operators $\Lambda_j^{(i)}$. Note that this protocol is equivalent to the two-qubit quantum state measurement protocol where each qubit is subjected to Pauli measurements. As this protocol is informationally complete\cite{bogdanov2011pra,bogdanov2010prl}, one can use it to reconstruct any two-qubit state or one-qubit process.

In order to form the $N$-qubit process measurement protocol, one could consider all the possible tensor products of vectors \eqref{eq:proto_prep} ($m_p = 6^N$) and operators \eqref{eq:proto_meas} ($m_m = 6^N$).

Note that one- and $N$-qubit protocols with cube symmetry form the decomposition of unity:
\begin{equation}\label{eq:proto_unity}
  \sum_{i=1}^{m_p}{\sum_{j=1}^{m_m}{\Lambda_j^{(i)}}} \sim I_{s^2}.
\end{equation}
The analysis of statistical properties of the protocols that form the decomposition of unity has been performed in\cite{bogdanov2013njp,bogdanov2011pra,bogdanov2010prl,rehacek2004,bogdanov2011jetp}.

\subsection{Reconstruction of $\chi$-matrix}\label{sect:chi_reconstruction}

In order to reconstruct $\chi$-matrix by measurement results we will use the maximum-likelihood estimation method together with the root approach\cite{bogdanov2011pra,bogdanov2009jetp,bogdanov2004optics}. Instead of determining the $\chi$-matrix we estimate its square root: matrix $e$ of dimension $s^2 \times r$ (see eq. \eqref{eq:chi_def}). This approach guarantees that the resulting $\chi$-matrix will be Hermitian and positive and lets us adjust its rank (i.e. the rank of the quantum process). Assume that all the measurements are being performed independently and that $t_j^{(i)}$ representatives of the $i$-th input state are used for the measurement of $\Lambda_j$. Then the probability to register $k_j^{(i)}$ events has the Poisson distribution with the expected value $t_j^{(i)} p_j^{(i)}$. The maximum-likelihood estimation of matrix $e$ in this case is reduced to numerical solution of the following quasi-linear equation (likelihood equation)\cite{bogdanov2011pra,bogdanov2010prl,bogdanov2009jetp,bogdanov2004pra}:
\begin{equation}
  Ie = J(e)e, \quad
  I = \sum_{i,j}{ t_j^{(i)}\Lambda_j^{(i)} }, \quad
  J(e) = \sum_{i,j}{ \frac{k_j^{(i)}}{p_j^{(i)}(ee^\dagger)}\Lambda_j^{(i)} }.
\end{equation}

Rank $r$ of the quantum process under consideration determines the dimension of matrix $e$. One can set it on the basis of an \textit{a priori} knowledge about the physics of the process, or by the adequate model constructing procedure (see Sect.~\ref{sect:chi_rank}).

To fulfill the normalization condition \eqref{eq:chi_normalization} one could complement the measurement statistics with a fictitious one that corresponds to the measurement of subsystem $A$\cite{bogdanov2013njp}. Let us consider the $\chi$-matrix measurement operator of the form $\Pi_\phi \otimes I_s$, where $\Pi_\phi$ is an arbitrary projector. If the $\chi$-matrix meet the condition \eqref{eq:chi_normalization} this measurement gives the probability (normalized to $s$)
\begin{equation}
  p_\phi(\chi) = \Tr(\chi(\Pi_\phi \otimes I_s)) = 1.
\end{equation}
Let us assume that this measurement is performed $t_\phi$ times, and use the expected value as the number of registered events ($k_\phi = t_\phi$). If the set of projectors $\Pi_\phi$ for different $\phi$ form an informationally complete protocol and $t_\phi \gg t_j^{(i)}$ for all $\phi$, $i$ and $j$, then such a complement of the measurement protocol will automatically make the reconstruction result to meet the condition \eqref{eq:chi_normalization}.

Let us denote the $\chi$-matrix of the quantum process under consideration by $\chi$ and the reconstruction result by $\hat\chi$. We will estimate the reconstruction accuracy as the fidelity between the corresponding Choi--Jamiolkowski state density matrices for $\chi$ and $\hat\chi$\cite{gilchrist2005}:
\begin{equation}
  F = \qty( \Tr\sqrt{\sqrt{\rho_\chi} \hat\rho_\chi \sqrt{\rho_\chi}} )^2, \quad \rho_\chi = \frac1s \chi, \quad \hat\rho_\chi = \frac1s \hat\chi.
\end{equation}

\subsection{Estimation of $\chi$-matrix rank}\label{sect:chi_rank}

To define the quantum process rank we go through its values and pick the value that adequately describes the statistical data. As the adequacy criterion we use the common chi-squared test\cite{bogdanov2009jetp}.

At first, let us consider the rank-1 model (this model corresponds to a unitary evolution) and perform the reconstruction according to Sect.~\ref{sect:chi_reconstruction}. The result of the reconstruction is the matrix $\hat\chi$ of rank 1. Next, let us calculate the following value:
\begin{equation}\label{eq:chi2}
  \chi_r^2 = \sum_{l=1}^m{ \frac{ (O_l-E_l)^2 }{E_l} },
\end{equation}
where $O_l$ are the numbers of registered events $k_j^{(i)}$ and $E_l$ are their expectations $t_j^{(i)} p_j^{(i)}(\hat\chi)$. It is known from classical statistics\cite{kendall1961} that the value of \eqref{eq:chi2} is the realization of the random variable $X$ that has the chi-squared distribution. The number of degrees of freedom in our case is equal to
\begin{equation}
  \nu = m - \nu_P - \nu_K,
\end{equation}
where $\nu_P = 2s^2r - r^2 - s^2$ is the number of real independent parameters which describe $\hat\chi$\cite{bogdanov2013njp} and $\nu_K$ is the number of relations in measurement results. In the case of completely independent measurements $\nu_K = 1$ (the only relation is set by the total number of measurements). If the measurement results are determined by the measurement of $m_b$ observables, then $\nu_K = m_b$ as the total probability of all the possible results for each observable is fixed and equal to 1.

After determining the number of degrees of freedom for $X$ one can calculate the probability that $X>\chi_r^2$. If this probability is less than the fixed significance level $\alpha$, then the rank-1 model is statistically insignificant and should be rejected. Then the above procedure is repeated with rank-2, rank-3, etc. (maximum rank-$s^2$) models until one gets a statistically significant result. If full-rank model ($r = s^2$) is also insignificant, then the instrumental errors prevail over statistical errors for a given sample size. In this case the rank of the quantum process could not be estimated by means of chi-squared test.

\section{FUZZY MEASUREMENTS APPROACH}\label{sect:fuzzy_measurements}

Let us perform the procedure of SPAM-errors consideration in the same way as it has been done for quantum states tomography in work\cite{bogdanov2016spie_fuzzy}. We will describe the measurements at the output of $\mathcal{E}$ by generalized positive operators of the form
\begin{equation}\label{eq:lam_fuzzy_general}
  \tilde\Lambda_j = \sum_k{f_k^j \ketbra{\varphi_k^j}},
\end{equation}
where each $f_k^j \geq 0$ sets the probability to measure the corresponding projector $\ketbra*{\varphi_k^j}$. When there is only one term in sum \eqref{eq:lam_fuzzy_general} such a measurement corresponds to projective measurement. The particular form of \eqref{eq:lam_fuzzy_general} depends on the physical measurement processes. The requirement for positivity of $\tilde\Lambda_j$ is set by the non-negativity of events probabilities.

Similarly, imperfections of the input state preparation procedure require using density matrices $\tilde\rho^{(i)}$ instead of $\ket*{\psi^{(i)}}$.

Thus, the measurement operators for $\chi$-matrix in the case of general SPAM-errors are the following:
\begin{equation}
  \tilde\Lambda_j^{(i)} = \qty(\tilde\rho^{(i)})^\ast \otimes \tilde\Lambda_j.
\end{equation}

\subsection{Noise-gate-noise (NGN) model}\label{sect:ngn_model}

\begin{figure} [ht]
  \centering
  \includegraphics[width=0.6\linewidth]{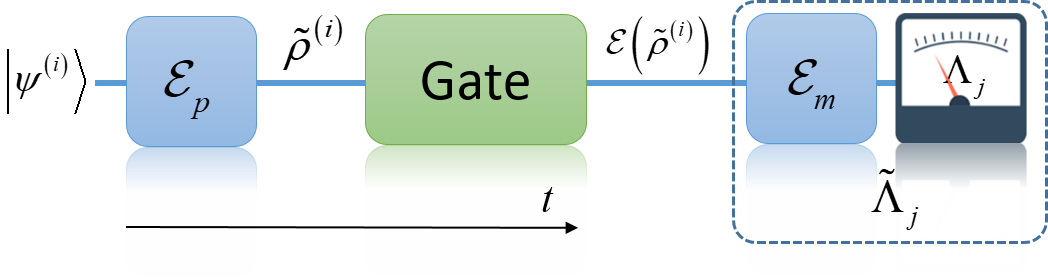}
  \caption{\label{fig:fuzzy_scheme_1} The scheme of imperfect measurements within NGN-model: the quantum state is subjected to errors after the perfect input state preparation and before the perfect measurement.}
\end{figure}

Let us consider a special case of SPAM-errors (Fig.~\ref{fig:fuzzy_scheme_1}). The quantum state which is perfectly prepared in $\ket*{\psi^{(i)}}$ is subjected to the rank-$r_p$ noisy quantum process $\mathcal{E}_p$ which is described with Kraus operators $E_{p,k}$ ($k = 1, 2, \dots, r_p$). At the output of the gate the state is subjected to the rank-$r_m$ noisy quantum process $\mathcal{E}_m$ with Kraus operators $E_{k,m}$ ($k = 1, 2, \dots, r_m$). Then for the measurement operator $\Lambda_j$ the probabilities \eqref{eq:dm_prop} turn out to be
\begin{equation}
  p_j^{(i)} = \Tr( \qty( \sum_{l=1}^{r_m}{ E_{m,l} \mathcal{E}\qty( \sum_{k=1}^{r_p}{ E_{p,k}\ketbra*{\psi^{(i)}}E_{p,k}^\dagger } ) E_{m,l}^\dagger } ) \Lambda_j ) = \Tr(\mathcal{E}(\tilde\rho^{(i)}) \tilde\Lambda_j),
\end{equation}
where the density matrix for the imperfectly prepared state and for operator for the imperfect measurement are
\begin{equation}\label{eq:fuzzy_ngn}
  \tilde\rho^{(i)} = \sum_{k=1}^{r_p}{ E_{p,k}\ketbra*{\psi^{(i)}}E_{p,k}^\dagger }, \quad
  \tilde\Lambda_j = \sum_{k=1}^{r_m}{ E_{m,k}^\dagger \Lambda_j E_{m,k} }.
\end{equation}
To get $\tilde\Lambda_j$ we have used the cyclic property of the trace.

Note that if the process $\mathcal{E}_m$ of measurement errors is trace preserving (see eq.~\eqref{eq:kraus_normalization} or eq.~\eqref{eq:chi_normalization}) then the protocol maintains the property of unity decomposition \eqref{eq:proto_unity}\cite{bogdanov2016spie_fuzzy}. However, this statement does not hold in general in the case of presence of preparation errors $\mathcal{E}_p$. This could be easily seen if $\mathcal{E}_p$ has the form of the energy damping channel. Under the impact of this channel any state gradually transforms to the ground state $\ket0$. In the limit of strong damping one would have $\tilde\rho^{(i)} \approx \ketbra0$ for each $i$, resulting in the violation of \eqref{eq:proto_unity}.

\subsection{Imperfect preparation and projective measurement (IPPM) model}\label{eq:ippm_model}

Let us now consider a more detailed $\chi$-matrix measurement model. This model consists of the following basic elements: the initialization of the state $\ket{\psi_0}$,  the preparation of the $i$-th input state by the unitary transformation $U_{prep}^{(i)}$ ($i = 1,2,\dots,m_p$), the performing of the transformation under consideration $\mathcal{E}$, the basis change by the unitary transformation $U_{basis}^{(l)}$ ($l = 1,2,\dots,m_b$), projective measurement in a fixed basis. The basis change gate is required for many physical systems where one can directly measure only a single observable (e.g. energy states population) with measurement operators $\Pi_k$ ($k = 1,2,\dots,s$) -- projectors to the eigenstates of this observable. If one preliminary subject the state to the transformation $U_{basis}^{(l)}$ then the subsequent projective measurement is equivalent to the measurement of the gate output state with operators $U_{basis}^{(l)\dagger} \Pi_k U_{basis}^{(l)}$. Similarly, usually one can experimentally initialize only a single quantum state $\ket{\psi_0}$ (e.g. by an optical pumping). Any other pure state could be obtained by the transformation of $\ket{\psi_0}$ with $U_{prep}^{(i)}$. To perform the $\chi$-matrix measurements of this type one should implement $m_p \cdot m_b$ quantum schemes. The total amount of measurement operators for $\chi$-matrix in this case is $m = m_p \cdot m_m$, where $m_m = s \cdot m_b$ is the amount of measurement operators for the output state of $\mathcal{E}$.

\begin{figure} [ht]
  \centering
  \includegraphics[width=0.8\linewidth]{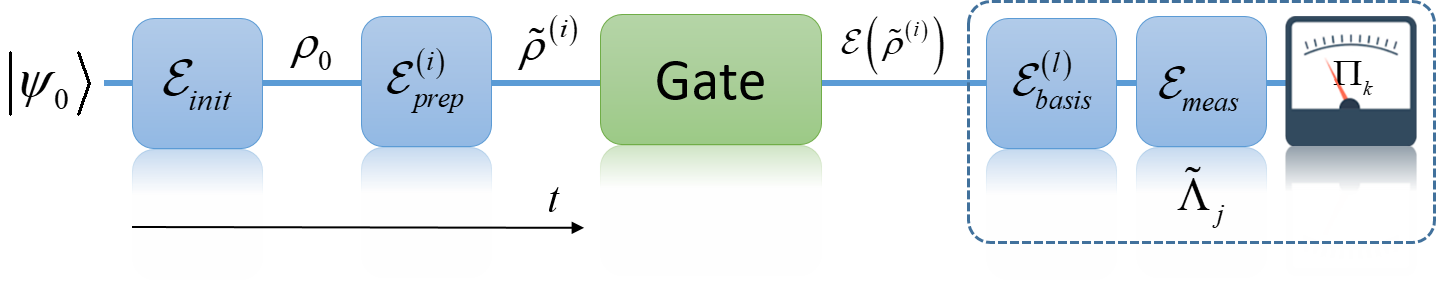}
  \caption{\label{fig:fuzzy_scheme_2} The scheme of imperfect measurements within IPPM-model: initialization and measurement are imperfect as well as preparation and basis change gates.}
\end{figure}

We will include SPAM-errors in this model in the following way (Fig.~\ref{fig:fuzzy_scheme_2}). The $l$-th basis change gate is described by a quantum process $\mathcal{E}_{basis}^{(l)}$ with Kraus operators $E_{basis,k}^{(l)}$ ($k = 1,2,\dots,r_{basis}^{(i)}$) instead of ideal unitary operator $U_{basis}^{(l)}$. We assume the imperfect measurements outcomes probabilities to be linear function of the measured density matrix elements so they can be described by a completely positive map $\mathcal{E}_{meas}$ (with Kraus operators $E_{meas,k}$, $k=1,2,\dots,r_{meas}$) followed by the ideal measurement device. Then for each basis change gate one can form $s$ positive measurement operators for the quantum state at the output of the gate under consideration:
\begin{equation}
  \tilde\Lambda_j \equiv \tilde\Lambda_{(l-1)s+k} = \sum_{m,n}{ E_{basis,n}^{(l)\dagger} E_{meas,m}^\dagger \Pi_k E_{meas,m}  E_{basis,n}^{(l)} }, \quad
  l = 1,2,\dots,m_b, \quad k = 1,2,\dots,s.
\end{equation}
Similarly, we assume that preparation gate is described with $\mathcal{E}_{prep}^{(i)}$ ($E_{prep,k}^{(i)}$, $k = 1,2,\dots,r_{prep}^{(i)}$) instead of $U_{prep}^{(i)}$ and that imperfect initialization could be expressed by the perfectly initialized $\ket{\psi_0}$ followed by $\mathcal{E}_{init}$ ($E_{init,k}$, $k=1,2,\dots,r_{init}$). Then
\begin{equation}
  \tilde\rho^{(i)} = \sum_{m,n}{ E_{prep,n}^{(i)} E_{init,m} \ketbra{\psi_0} E_{init,m}^\dagger E_{prep,n}^{(i)\dagger} }, \quad
  i = 1,2,\dots,m_p.
\end{equation}

Consider the qubit initialization in the state
\begin{equation}
  \rho_0 = p\ketbra{\psi_0} + (1-p)\ketbra{\psi_1},
\end{equation}
where $1/2 < p \leq 1$ and $\braket{\psi_1}{\psi_0}=0$. Then one can describe $\mathcal{E}_{init}$ in a form of the depolarizing noise with the probability of depolarization $\gamma = 2(1-p)$:
\begin{equation}\label{eq:init_depol}
  \rho_0 = \mathcal{E}_{init}(\ketbra{\psi_0}) = \mathcal{E}_{\gamma}(\ketbra{\psi_0}), \quad
  \mathcal{E}_{\gamma}(\rho) = (1-\gamma)\rho + \frac\gamma2 I_2.
\end{equation}
As the depolarizing noise commutes with any transformation, one could identify $\mathcal{E}_{init}$ as the qubit measurement error. However, this is generally not true for a multi-qubit register where the local depolarizing noise influences each qubit separately.

Note that NGN-model (Sect.~\ref{sect:ngn_model}) is the special case of a more general IPPM-model, described in this section, if one takes projectors to eigenvectors of a set of observables as the measurement operators. Then $\mathcal{E}_{init}$ and $\mathcal{E}_{meas}$ are identity transformations and
\begin{equation}
  E_{prep,k}^{(i)} = E_{p,k}U_{prep}^{(i)}, \quad E_{basis,k}^{(l)} = U_{basis}^{(l)}E_{m,k}
\end{equation}
for each $i$, $l$ and $k$.

\subsection{Example of QPT with fuzzy measurements protocol}\label{sect:ippm_example}

Let us consider the example of one-qubit gate tomography within the IPPM-model. We assume that the measurement device performs the measurement of the Pauli observable $\sigma_z$ and that $\ket{\psi_0} = \ket0$ ($\ket0$ and $\ket1$ are eigenstates of Pauli matrix $\sigma_z$). To implement the cube protocol (Sect.~\ref{sect:chi_measurement}) one should be able to do one-qubit rotations described by the unitary operator
\begin{equation}\label{eq:1q_rotation}
  R_{\va{n}}(\theta) = \exp(-i \frac\theta2 (n_x\sigma_x + n_y\sigma_y + n_z\sigma_z)), \quad \theta,n_x,n_y,n_z \in \mathbb{R}, \quad \abs{\va{n}} = 1.
\end{equation}
The values of $\theta$ and $\va{n}$ for cube protocol are listed in Table~\ref{tab:proto_cube}.

\begin{table}[ht]
  \caption{Parameters of one-qubit rotations for preparing 6 input states \eqref{eq:proto_prep} for cube protocol. To measure $\sigma_z$, $\sigma_y$ and $\sigma_z$ observables one could consider rotations with parameters 1, 3 and 5 respectively. The case $\theta = 0$ corresponds to the absence of transformation.}
  \label{tab:proto_cube}
  \begin{center}
    \renewcommand{\arraystretch}{1.2}
    \begin{tabular}{|C{.8cm}|C{1.5cm}|C{1.5cm}|C{1.5cm}|C{1.5cm}|C{1.5cm}|C{1.5cm}|}
      \hline
       & \textbf{1} & \textbf{2} & \textbf{3} & \textbf{4} & \textbf{5} & \textbf{6} \\
      \hline
      $\theta$ & $\pi$ & $\pi$ & $\pi$ & $\pi$ & $0$ & $\pi$ \\
      \hline
      $n_x$ & $1/\sqrt2$ & $-1/\sqrt2$ & $0$ & $0$ & -- & $1$ \\
      \hline
      $n_y$ & $0$ & $0$ & $1/\sqrt2$ & $-1/\sqrt2$ & -- & $0$ \\
      \hline
      $n_z$ & $1/\sqrt2$ & $1/\sqrt2$ & $1/\sqrt2$ & $1/\sqrt2$ & -- & $0$ \\
      \hline
    \end{tabular}
    \renewcommand{\arraystretch}{1}
  \end{center}
\end{table}  

Assume that $\mathcal{E}_{init}$ is the depolarizing noise \eqref{eq:init_depol} with $\gamma = 0.01$. Preparation and basis change gates are being performed together with the impact of amplitude damping and dephasing with parameters $T_1 = 100T$ and $T_2 = 50T$ where $T$ is the time duration of the gate (we have performed the simulation of these transformation on the basis of the works\cite{bogdanov2013spie_noise,bogdanov2014spie_dephasing}). Finally, we assume that $\mathcal{E}_{meas}$ is the amplitude damping channel of unit time duration with $T_1 = 20$. Such a measurement error corresponds to the increasing of the probability to get $\ket0$ and decreasing of the probability to get $\ket1$ in comparison to the ideal case.

\begin{figure}[ht]
  \begin{center}
    \begin{minipage}[ht]{0.325\linewidth}\centering
      \includegraphics[width=\linewidth]{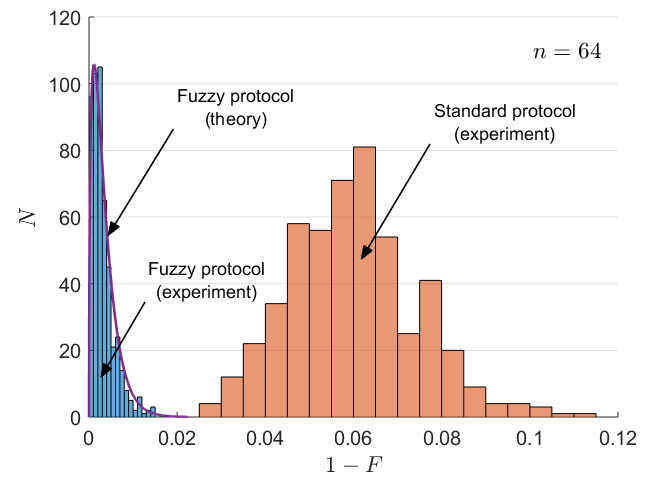} \\ a
    \end{minipage}
    \hfill
    \begin{minipage}[ht]{0.325\linewidth}\centering
      \includegraphics[width=\linewidth]{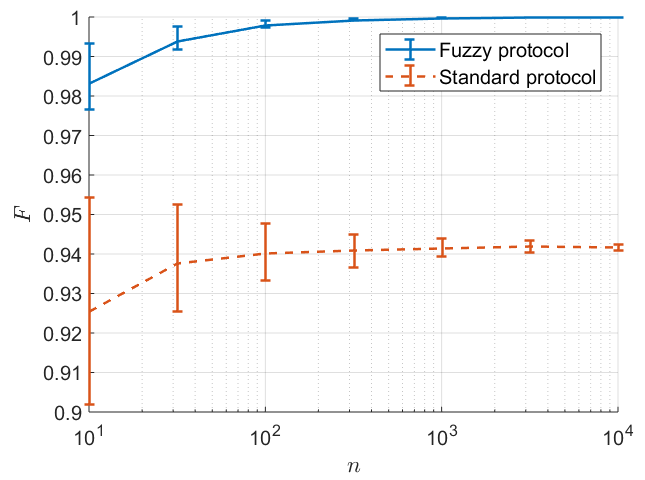} \\ b
    \end{minipage}
    \hfill
    \begin{minipage}[ht]{0.325\linewidth}\centering
      \includegraphics[width=\linewidth]{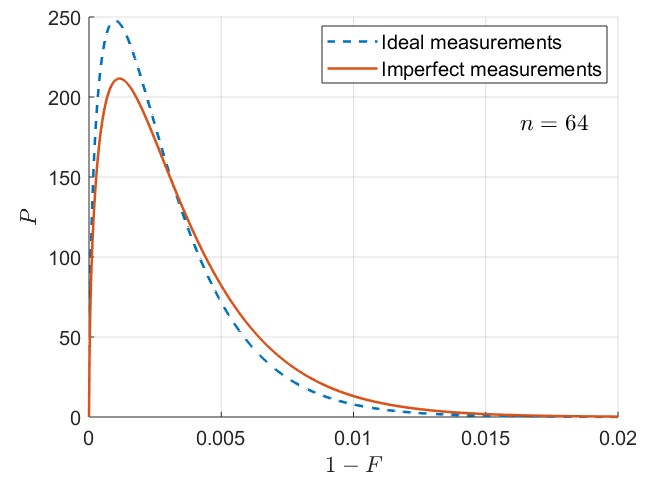} \\ c
    \end{minipage}
  \end{center}
  \caption{\label{fig:res_ippm} Hadamard gate tomography. (a) Infidelity distribution: histograms are the results over 500 numerical experiments, solid line is the theoretical distribution. (b) Fidelity vs. sample size: for each $n$ the mean fidelity and confidence interval (the first and third quantiles) over 500 numerical experiments were calculated. (c) Theoretical distributions for ideal (no SPAM-errors) and imperfect measurements.}
\end{figure}

Fig.~\ref{fig:res_ippm} depicts the simulation results for QPT of the unitary ($r = 1$) Hadamard gate. The statistical data was obtained by Monte Carlo method. We have also obtained the theoretical fidelity distribution based on the complete information matrix and the universal fidelity distribution\cite{bogdanov2013njp,bogdanov2010prl,bogdanov2009jetp}. The fuzzy measurements protocol includes all the considered SPAM-errors and provides an asymptotically optimal estimation of the quantum process while the standard protocol that considers ideal SPAM only gives the biased estimation and the limited reconstruction fidelity.

For the fuzzy measurements protocol the rank-1 model turned out to be the significant one. This is consistent with the true gate rank $r = 1$. For the standard protocol the rank-3 model was significant. That means that SPAM-errors introduced additional non-unitary components to the results of process reconstruction.

However, one should also note that SPAM-errors result in the loss of information about the $\chi$-matrix parameters. This is demonstrated in Fig.~\ref{fig:res_ippm}c where the theoretical infidelity distributions for the perfect measurements (no SPAM-errors) and imperfect ones are presented.

Note that fuzzy measurements protocol could be only implemented if one knows all the SPAM-errors characteristics with a high accuracy. One could try to measure these characteristics using QPT but it would be also subjected to the same SPAM-errors. Below we will show how one can estimate part of the SPAM-errors and increase the fidelity of further measurements.

\section{TOMOGRAPHY OF MEASUREMENT IMPERFECTION}\label{sect:empty_tomo}

To define SPAM-errors experimentally one could measure a known process. The ``empty'' gate -- transformation that acts during time $t=0$ (i.e. the absence of a gate) -- performs the ideal identity transformation and is known precisely. In this case one should perform the measurement right after the state preparation. The result of the ``empty'' gate tomography would be some $\chi$-matrix that contains part of the SPAM-errors. One could then calculate Kraus matrices that correspond to this $\chi$-matrix and form the fuzzy measurements protocol using GN (gate-noise) model where one replaces $\Lambda_j$ with $\tilde\Lambda_j$ according to \eqref{eq:fuzzy_ngn} leaving $\ket*{\psi^{(i)}}$ unchanged. The accuracy of this approach depends on the commutation relations between SPAM-errors and the gates that one uses during QPT.

If it is known that preparation errors prevail over measurement errors, one should use ``empty'' gate $\chi$-matrix within the NG (noise-gate) model: one adjusts $\ket*{\psi^{(i)}}$ according to \eqref{eq:fuzzy_ngn} leaving $\Lambda_j$ unchanged.

\begin{figure} [ht]
  \centering
  \includegraphics[width=0.35\linewidth]{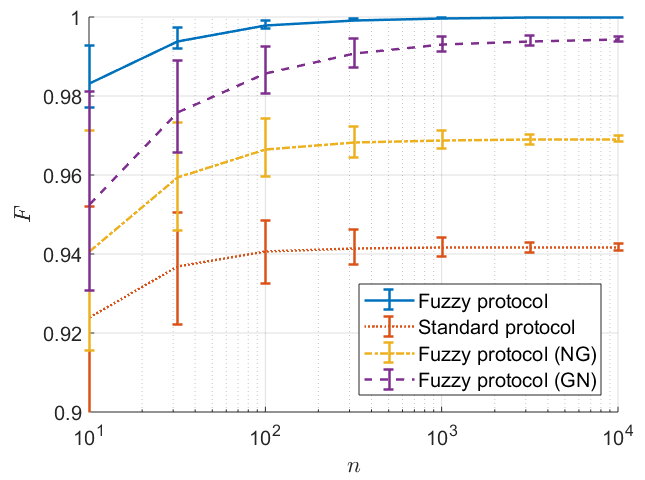}
  \caption{\label{fig:res_ngn} Fidelity of Hadamard gate tomography with different protocols vs. sample size:  for each $n$ the mean fidelity and confidence interval (the first and third quantiles) over 500 numerical experiments were calculated.}
\end{figure}

We have performed the simulation of the above approach with SPAM-errors described in \ref{sect:ippm_example}. To measure ``empty'' gate we have used $n = 1000$ copies of the each input state. The resulting $\chi$-matrix have been used for further tomography with use of fuzzy measurements protocol within GN- and NG- models. The results for ideal Hadamard gate are presented in Fig.~\ref{fig:res_ngn}. Both GN- and NG-model provide higher fidelity and give rank-2 $\chi$-matrix as significant one. The results mean that we obtained only a partial information about the SPAM-errors by the ``empty'' gate tomography in these conditions.

Note that GN-model turned out to be more accurate than NG-model. This could be explained by the fact that we have taken the depolarizing noise at initialization. This noise commutes with any transformation and could be considered as preparation noise as well as measurement noise. At the same time the measurement noise in the form of amplitude damping channel does not commute with some of the basis change gates and with the Hadamard gate itself. This makes the quality of the NG-model much lower.

\section{EXPERIMENT: IBM Q}\label{sect:ibm}

We have tested the above approach experimentally using the superconducting 5-qubit quantum processor IBM~Q~5 Tenerife (ibmqx4) which is available through the cloud service\cite{ibm2017}.

We used the cube protocol and in each experiment the number of the gate input states representatives was $n = 1000$. At first, we performed the tomography of the ``empty'' gate for qubits Q0 and Q1 using standard protocol. The fidelity of the resulting $\chi$-matrices in comparison to the ideal identity gate was 89.81\% for Q0 and 88.19\% for Q1. For both qubits the rank-3 model was statistically significant. Note that this is important to measure SPAM-errors for each qubit as they may differ significantly.

\begin{figure}[ht]
  \begin{center}
    \begin{tabular}{ccc}
      \includegraphics[width=0.2\linewidth]{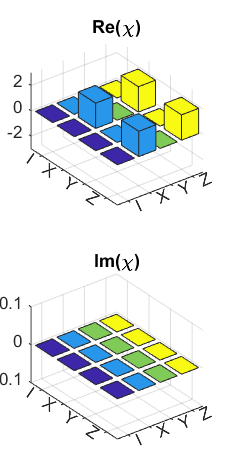} &
      \includegraphics[width=0.2\linewidth]{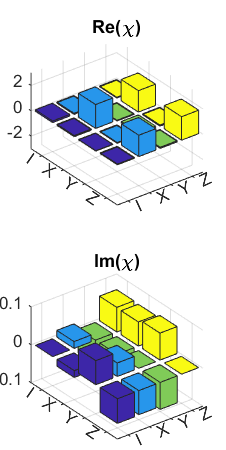} &
      \includegraphics[width=0.2\linewidth]{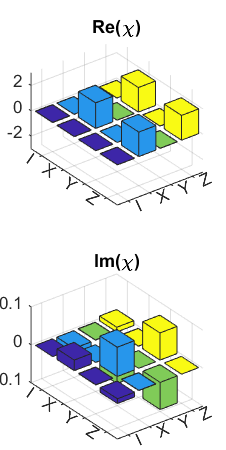} \\
      a & b & c
    \end{tabular}
  \end{center}
  \caption{\label{fig:res_ibm_h} Real and imaginary parts of the Hadamard gate $\chi$-matrices in Pauli representation: ideal (a), reconstructed with standard (b) and GN-model fuzzy (c) protocols.}
\end{figure}

\begin{figure}[ht]
  \begin{center}
    \includegraphics[width=0.55\linewidth]{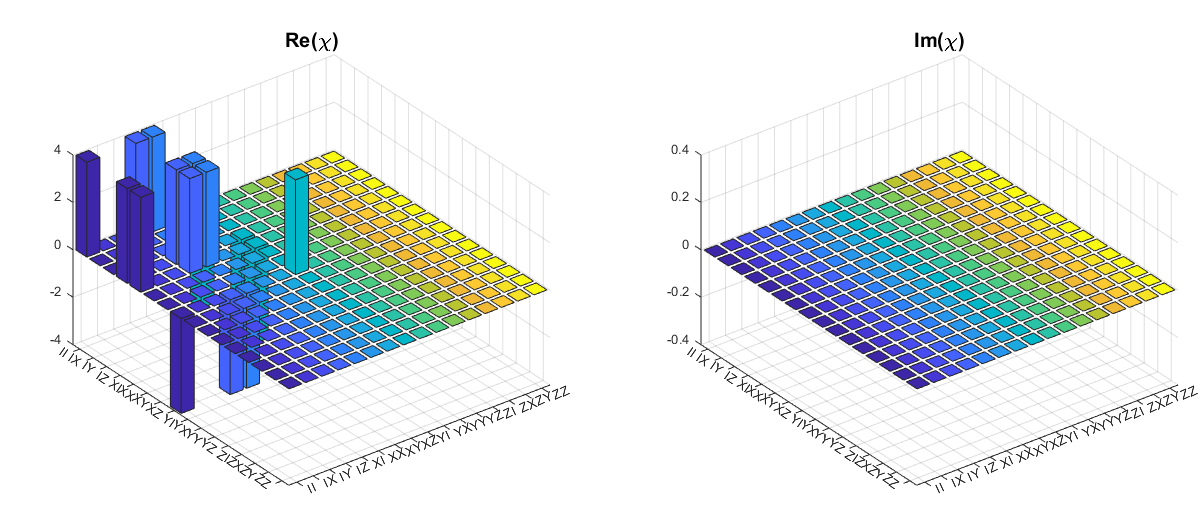} \\ a \\
    \includegraphics[width=0.55\linewidth]{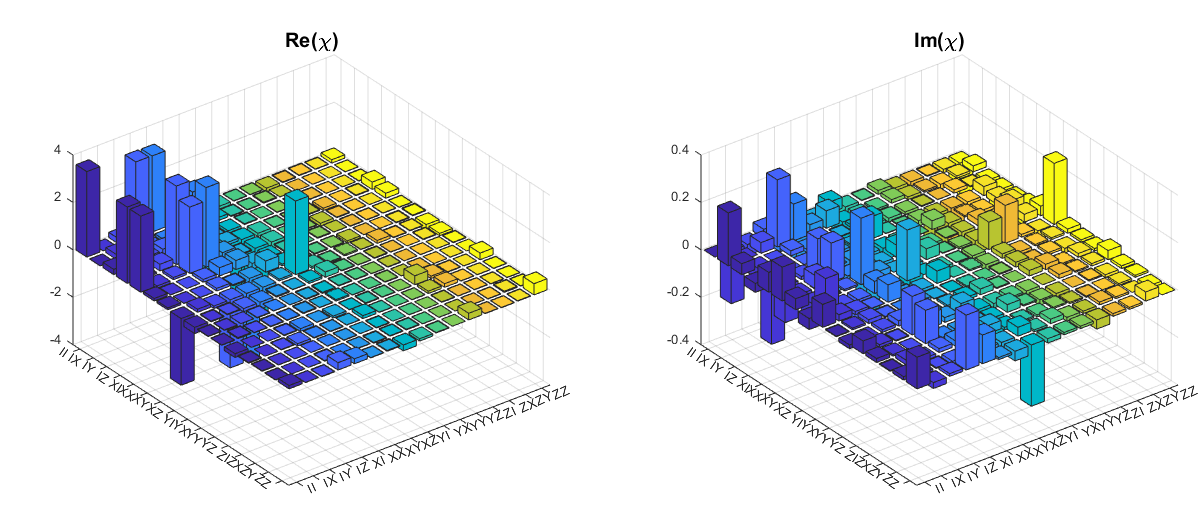} \\ b \\
    \includegraphics[width=0.55\linewidth]{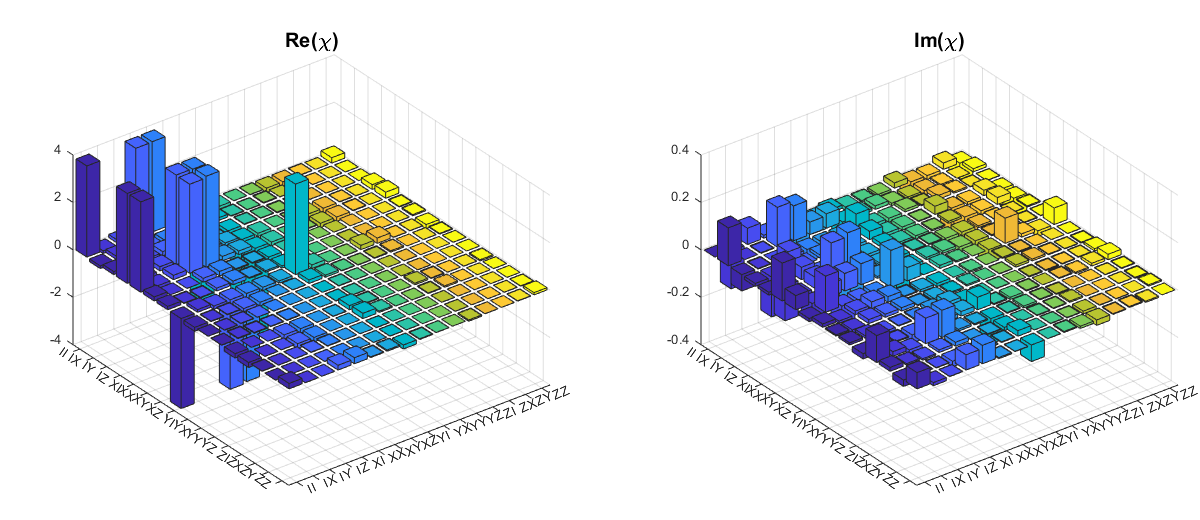} \\ c
  \end{center}
  \caption{\label{fig:res_ibm_cnot} Real and imaginary parts of the CNOT gate $\chi$-matrices in Pauli representation: ideal (a), reconstructed with standard (b) and GN-model fuzzy (c) protocols.}
\end{figure}

Next, we performed the tomography of Hadamard (Fig.~\ref{fig:res_ibm_h}) and CNOT (Fig.~\ref{fig:res_ibm_cnot}) gates using standard and fuzzy protocols but with the same statistical data. To depict $\chi$-matrices we used Pauli representation: instead of \eqref{eq:chi_basis} we use
\begin{equation}
  A_0 = I = \frac{1}{\sqrt2}I_2,\quad
  A_1 = X = \frac{1}{\sqrt2}\sigma_x,\quad
  A_2 = Y = \frac{1}{\sqrt2}\sigma_y,\quad
  A_3 = Z = \frac{1}{\sqrt2}\sigma_z.\quad
\end{equation}
For two-qubit gates one has to consider all 16 combinations of $I$, $X$, $Y$ and $Z$.

Results of reconstruction are presented in Table.~\ref{tab:ibm_results}. Both GN- and NG-model fuzzy measurements protocols give higher reconstruction fidelity than the standard protocol which considers ideal measurements only. With GN-model we improved the reconstruction fidelity from 89\% to 98\% for Hadamard gate and from 77\% to 95\% for CNOT gate. One can also observe the decreasing of the gates rank.

Note that the values of fidelity in Table~\ref{tab:ibm_results} were calculated in relation to ideal transformations. As the real transformations are probably not ideal the values of $1-F$ contain both reconstruction errors and gates implementation errors. The results for standard protocol are in a close agreement with the results obtained in the work\cite{shukla2018}, where the reconstruction was performed by linear inversion.

\begin{table}[ht]
  \caption{The results of the QPT on IBM quantum processor for different protocols. Sign ``--'' for $r$ means that instrumental errors were too high to estimate the process rank.}
  \label{tab:ibm_results}
  \begin{center}
    \renewcommand{\arraystretch}{1.2}
    \begin{tabular}{|l|C{1.5cm}|C{1.5cm}|C{1.5cm}|C{1.5cm}|}
      \hline
      \multirow{2}{*}{} & \multicolumn{2}{c|}{Hadamard (Q0)} & \multicolumn{2}{c|}{CNOT (Q1 $\rightarrow$ Q0)} \\
      \cline{2-5}
       & $r$ & $F$, \% & $r$ & $F$, \% \\
      \hlineB{2.5}
      \textbf{Standard protocol} & 3 & 89.02 & 8 & 76.50 \\
      \hline
      \textbf{Fuzzy protocol (GN-model)} & 2 & \textbf{98.13} & 5 & \textbf{94.66} \\
      \hline
      \textbf{Fuzzy protocol (NG-model)} & -- & 95.50 & -- & 88.29 \\
      \hline
    \end{tabular}
    \renewcommand{\arraystretch}{1}
  \end{center}
\end{table}

It is worth mentioning that SPAM-errors have to be homogeneous in time in order to perform the above procedure. If preparation and measurement parameters are changing significantly within the time period between the ``empty'' gate tomography and tomography of the gate under consideration, the fuzzy measurements protocol will not adequately describe the measurement process.

\section{CONCLUSIONS}\label{sect:conclusions}

The presence of SPAM-errors significantly limits the quality of QPT. To solve this problem, one should develop reconstruction methods that could effectively consider SPAM-errors and provide an adequate and consistent estimation of the process $\chi$-matrix. In the current work we have proposed the fuzzy measurements approach which gives an asymptotically optimal estimation when one has detailed information about the errors.

This information could in part be obtained experimentally by performing the tomography of the ideal identity gate. This gate could be implemented by performing the measurement right after the initial state preparation. The result of the reconstruction in this case is the $\chi$-matrix that contains partial information about the SPAM-errors. We have shown that this $\chi$-matrix could be used to significantly improve further tomography procedures.

This approach has been tested experimentally on the IBM superconducting quantum processor. As a result, the reconstruction fidelity has increased from 89\% to 98\% for Hadamard gate and from 77\% to 95\% for CNOT gate.

The results obtained in this work are of practical importance for the provision of the quality of quantum computation element base.

\acknowledgments % equivalent to \section*{ACKNOWLEDGMENTS}

This work was supported by Russian Foundation of Basic Research (project 18-37-00204).

% References
\bibliography{QI2018_FuzzyIBM_en} % bibliography data in QI2018_FuzzyIBM_en.bib
\bibliographystyle{spiebib} % makes bibtex use spiebib.bst

\end{document}